\documentclass[3p]{elsarticle}

\usepackage{todonotes}
\usepackage[export]{adjustbox} 
\usepackage{amsmath}
\usepackage{xcolor} 
\usepackage{subcaption} 
\usepackage{hyperref} 

\providecommand{\norm}[1]{\lVert#1\rVert}

\graphicspath{{"./figures/"}}

\usepackage{lineno}

\begin{document}

\title{Maintaining the proliferative cell niche in multicellular models of epithelia}

\author[1,2]{Claire Miller} \ead{c.m.miller@uva.nl}
\author[2,3,4]{Edmund Crampin} \ead{edmund.crampin@unimelb.edu.au}
\author[1]{James M. Osborne\corref{cor1}} \ead{jmosborne@unimelb.edu.au}

\cortext[cor1]{Corresponding author}

\address[1]{School of Mathematics and Statistics, The University of Melbourne, Parkville, Victoria 3010, Australia}
\address[2]{Systems Biology Laboratory, School of Mathematics and Statistics, and Department of Biomedical Engineering, The University of Melbourne, Parkville, Victoria 3010, Australia}
\address[3]{School of Medicine, Faculty of Medicine, Dentistry and Health Sciences, The University of Melbourne, Parkville, Victoria 3010}
\address[4]{ARC Centre of Excellence in Convergent Bio-Nano Science and Technology, Melbourne School of Engineering, The University of Melbourne, Parkville, Victoria 3010, Australia}

\begin{abstract}
The maintenance of the proliferative cell niche is critical to epithelial tissue morphology and function.
In this paper we investigate how current modelling methods can result in the erroneous loss of proliferative cells from the proliferative cell niche. 
Using an established model of the inter-follicular epidermis we find there is a limit to the proliferative cell densities that can be maintained in the basal layer (the niche) if we do not include additional mechanisms to stop the loss of proliferative cells from the niche. 
We suggest a new methodology that enables maintenance of a desired homeostatic population of proliferative cells in the niche: a rotational force is applied to the two daughter cells during the mitotic phase of division to enforce a particular division direction. 
We demonstrate that this new methodology achieves this goal. 
This methodology reflects the regulation of the orientation of cell division. 
\end{abstract}

\maketitle

\subsection*{Keywords:} epidermis; mitosis; agent-based models; computational biology

\section{Introduction}

Epithelial tissues, such as the epidermis or the colonic crypt, are commonly studied \textit{in silico} using multicellular, or agent based, models \cite{adra10,du18,li13,meineke01,mirams12,sutterlin17}. 
A popular subset of these models are cell centre off-lattice mechanics based methods which model cells individually (as points in space) and represent cell-cell interactions with forces, without restricting cell locations to a grid. 
An often used method for modelling cell division in cell centre models splits the parent cell into a pair of daughter cells which experience different intercellular forces to a normal pair of cells \cite{drasdo03,meineke01,mirams12,pitt-francis09}.
This method is intended to represent cell elongation and splitting during the final phase of the cell division process. 
However, we will show that, without including additional measures, such as the pinning of cells or additional forces, this method cannot maintain proliferative cell populations above certain, potentially biologically unrealistic, densities. 

In this paper we focus on the interfollicular epidermis (IFE), the outermost layer of the skin, though the results from this paper have implications to all epithelial tissues. 
The epidermis is the most superficial tissue in the skin and is the interface between our body and its external environment.
It has a distinct structure with a single layer at the base on an undulating basal membrane containing the proliferative cells, called the basal layer, which sits beneath layers of terminally differentiated (non-proliferative) cells. 
The basal layer is where all proliferation occurs in the IFE by means of one or two proliferative cell types \cite{alcolea13,clayton07,kaur11,mascre12,sada16}, making it the proliferative cell niche for the tissue.
As there is uncertainty around the proliferative cell types in the epidermis, in this paper we use the generic term proliferative cell rather than specifically defining a cell lineage.
We define niche only as the location in which proliferative cells are found in the tissue, without implying any specific form of extrinsic regulation of the cells.
Additionally, in the case of the epidermis, this niche could contain stem, progenitor, and differentiated cells \cite{kaur00,mascre12,sada16}.

\begin{figure} \centering
	\includegraphics[width=0.8\linewidth]{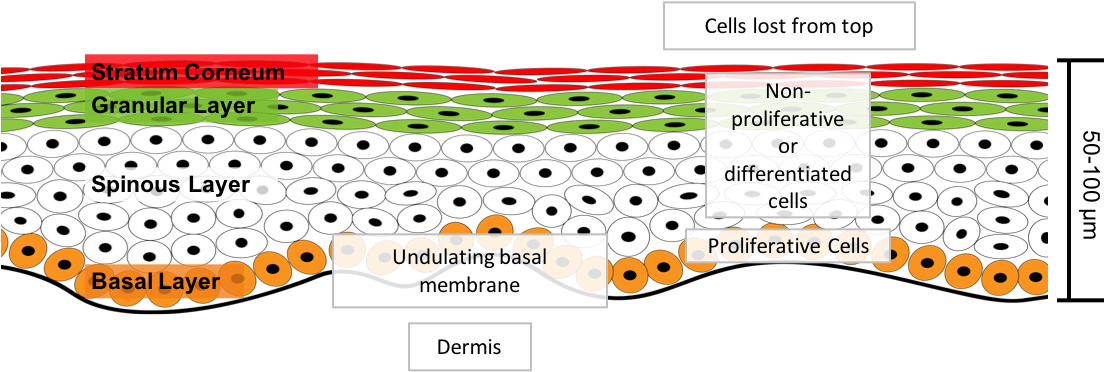}
	\caption{The structure of the inter-follicular epidermis (IFE). This diagram illustrates the different layers and some of their properties.}
	\label{fig:ifestructure}
\end{figure}

The density of proliferative cells in the basal layer of the IFE is not known.
Estimates of growth fraction (proportion of basal cells that are actively cycling) have varied from 20\% \cite{heenen98} to 60\% \cite{weinstein84} in biopsies of normal human epidermis.
The continuous proliferation in the basal layer ensures a steady replacement of the cells lost at the top of the tissue.
Consequently, maintaining the population of proliferative cells is critical to homeostasis of the tissue.
Cells in the IFE are removed (sloughed) from the top of the tissue and lost to the environment.
A representative diagram of the tissue is shown in Figure~\ref{fig:ifestructure}. 
Within humans, the IFE is approximately 50--100~$\mu m$ deep and is fully renewed every four weeks \cite{cursons15,menon12}.
The IFE consists of 95\% keratinocytes \cite{menon02} which govern the majority of the epidermal structure and dynamics. 
Consequently, we only model keratinocytes in this paper.

Recent work has used multicellular models to investigate the dynamics of the IFE \cite{li13,sutterlin17}. 
These models were used to investigate the morphology of the IFE \cite{sutterlin17} and clonal dynamics, that is the sizes of the cell populations stemming from a single proliferative cell \cite{li13}. 
Both of these tissue properties are influenced by the proliferative cell density. 
The morphology of proliferative and non-proliferative regions changes with the loss of proliferative cells, as could the stratification of the layers as the vertical velocities of cells are affected by reduced basal divisions. 
If the modelling techniques we use are influencing the proliferative cell population they would also influence the clonal dynamics, likely decreasing the number of clonal populations observed in the system. 
This also applies to clonal dynamics studies using these techniques in other epithelial tissues, such as the colonic crypt (intestinal glands in the colon) \cite{mirams12}.

\begin{figure}
    \centering
    \includegraphics[width=\linewidth]{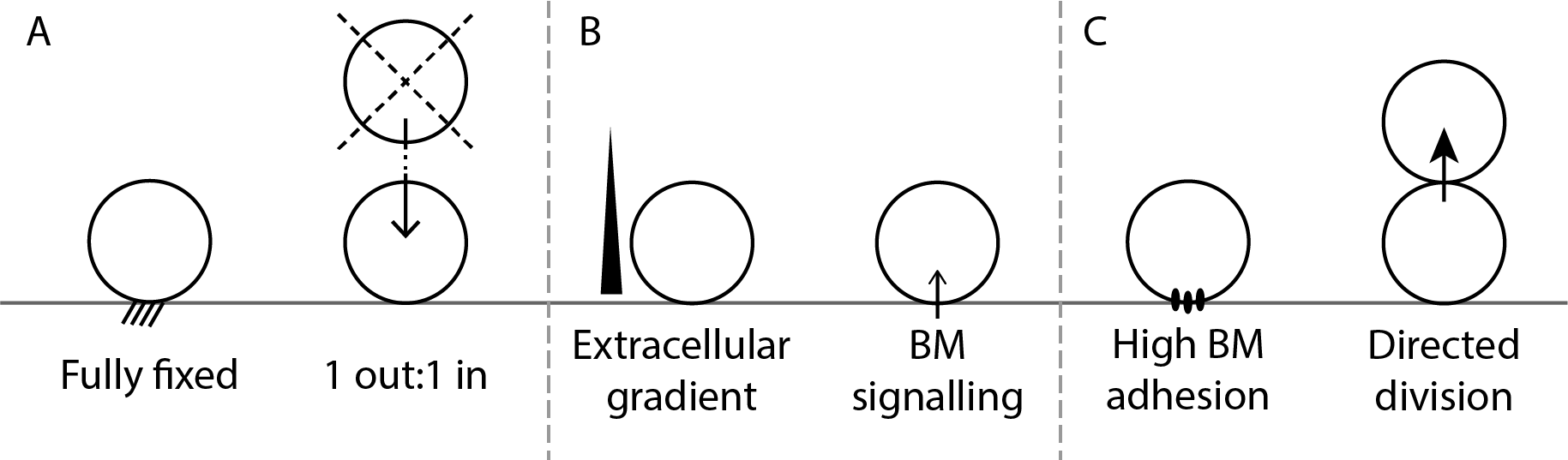}
    \caption{
    Potential methods for maintenance of a proliferative niche. 
    A: examples of previously used methods which maintain exact population numbers but are unlikely (1 out:1 in: a cell death triggers division of a basal cell).
    B: examples of signalling methods which guarantee proliferative niche control but require additional mechanisms to control for excessive proliferation.
    C: examples of methods which start with a number of proliferative cells and aim to maintain that number.
    In reality it is likely a combination of multiple methods. 
    Directed division is the method we explore in this paper. 
    BM: basal membrane.}
    \label{fig:proliferative_niche}
\end{figure}

There are several different strategies we know of to maintain proliferative cell populations.
These are shown in Figure~\ref{fig:proliferative_niche}.
It is likely that, biologically, several of these methods are occurring and contributing to niche maintenance.

The first panel, A, in Figure~\ref{fig:proliferative_niche}, shows methods for which there is no \textit{in vivo} or \textit{in vitro} evidence that we are aware of.
The first of these; cells pinned to the membrane, could be considered an extreme version of high membrane adhesion, shown in Panel C.
This pinning approach has been used previously in agent based models of the epidermis, such as in \citet{sutterlin17}.
Though this does guarantee maintenance of the niche, it is not biologically realistic and would be unable to support population asymmetry lineages, as symmetric divisions into the basal layer would require a rearrangement of basal cells to accommodate the new cell.
From a modelling standpoint, one would expect such an approach to be reasonable when analysing system dynamics in the higher layers of the tissue, however it would be expected to affect cell packing, and hence dynamics, in the lower layers.

The second approach in Panel A of Figure~\ref{fig:proliferative_niche} is the 1 out: 1 in policy.
In this approach a cell loss in the tissue triggers a proliferation event somewhere in the basal layer. 
This assumes all cells in the basal layer have proliferative potential, similar to the approaches in Panel B, however the key difference is that proliferation is triggered by long distance signalling from the top of the tissue, rather than from a pre-determined cycle length.
The inverted approach, 1 in: 1 out, has been used previously for colonic crypts in \citet{germano21}.
Though this maintains proliferation rates, it does not explicitly maintain a proliferative population in the basal layer.
Using such a system, it would be difficult to reproduce population asymmetry dynamics, or more generally any lineage dynamics involving multiple proliferative cell types and division policies.
Additionally, such a precise approach to tissue population size maintenance is highly unlikely in a biological system.

The second panel (B) in Figure~\ref{fig:proliferative_niche} shows two biochemical approaches that involve extracellular signalling to control division.
These approaches associate proliferative ability to the spatial location of a cell. 
This concept is commonly used when modelling the colonic crypt \cite{buske11,mirams12,vanleeuwen09}. 
In the crypt, cells proliferate at the base and become terminally differentiated towards the top of the tissue. 
This is thought to be due to a gradient of the signalling factor Wnt through the depth of the crypt and local Notch signalling \cite{vanleeuwen09,buske14}, and potentially a coupling between the niche and local tissue curvature via specialised cells (Paneth cells) \cite{buske12,buske14}. 
Using a signalling approach the niche will always be maintained. 
Signalling, from the membrane or extracellular space, will of course play a significant role in proliferative cell dynamics in the basal cell layer---a potential role for it in directed division (Panel C in the figure) is discussed later---but we argue this role may not be as the primary determinant of cell type.
This is because, as mentioned above, it has been hypothesised that two proliferative cell types exist in the IFE \cite{clayton07,mascre12,sada16}. 
In addition, in the IFE differentiated cells are also thought to reside in the basal layer \cite{kaur00}.
This suggests contact with the membrane, or an extracellular gradient, does not cause a cell to be of a certain type, and instead implies it could be an inherent trait, or more complicated intra- and inter- cellular signalling is required to maintain the niche.
There is however evidence that detachment from the basal membrane into suprabasal layers triggers differentiation in monolayers and developing tissue \cite{miroshnikova18}.
This may also be a factor in adult tissue, in addition to the maintenance mechanism(s) of the tissue, if any detachment of proliferative cells occurs.

The final panel, (C), in Figure~\ref{fig:proliferative_niche} shows mechanistic methods for maintenance of the proliferative niche which are biologically realistic.
The first is a commonly used method, and likely one contributing factor: high adhesion with the membrane.
It is known that the adhesion molecules between basal cells and the basal membrane are different than adhesion molecules between cells (hemidesmosomes compared to desmosomes) \cite{alberts15,tervonen11}.
Two agent based models of the epidermis, \citet{li13} and \citet{kobayashi16}, tightly bind proliferative cells to the membrane, with a dependence on proliferative cell type, by increasing adhesion.
This does help mitigate proliferative cell loss, however there is no evidence to show that different cell types have different adhesion levels to the membrane, and how strong any proliferative cell-membrane adhesion would be compared to other adhesion strengths.
Additionally, this method restricts proliferative cell movement which may not be biologically realistic and does not always guarantee maintenance of the population.

Using a multicellular model, which we will describe below, with membrane adhesion functions and parameters based on those used in \citet{li13}, we found the system could not maintain high populations of proliferative cells. 
The desired system structure and a representative example of a simulation result is shown in Figure~\ref{fig:simulationoutput}, and a video of the simulation result is shown in Supplementary Movie~1. 
It can be seen that almost all of the proliferative cells (red) have been lost from the basal layer and are moving upwards through the tissue. 
Figure~\ref{fig:cellloss} shows the proportion of original proliferative cells in the basal layer over time for 25~realisations of the model. 
The graphs shows that proliferative cell populations in the niche reduced to less than 20\% of their original population within 15~days. 
The mean proliferative cell count after 1,000~days is 2.8~cells, and 56\% of the simulations have no proliferative cells remaining.
Assuming a cell diameter of $10~\mu$m, this is a mean density of 2.8~proliferative cells/$10^4~\mu$m$^2$, and 56\% of simulations have a density of 0~proliferative cells/$10^4~\mu$m$^2$.  
The model description for these results is discussed in detail in the model section.

\begin{figure} \centering
	\begin{subfigure}{0.8\linewidth}
		\includegraphics[width=\linewidth]{simulation_output}
		\caption{}
		\label{fig:simulationoutput}
	\end{subfigure}
	\begin{subfigure}{0.4\linewidth}
		\includegraphics[width=\linewidth]{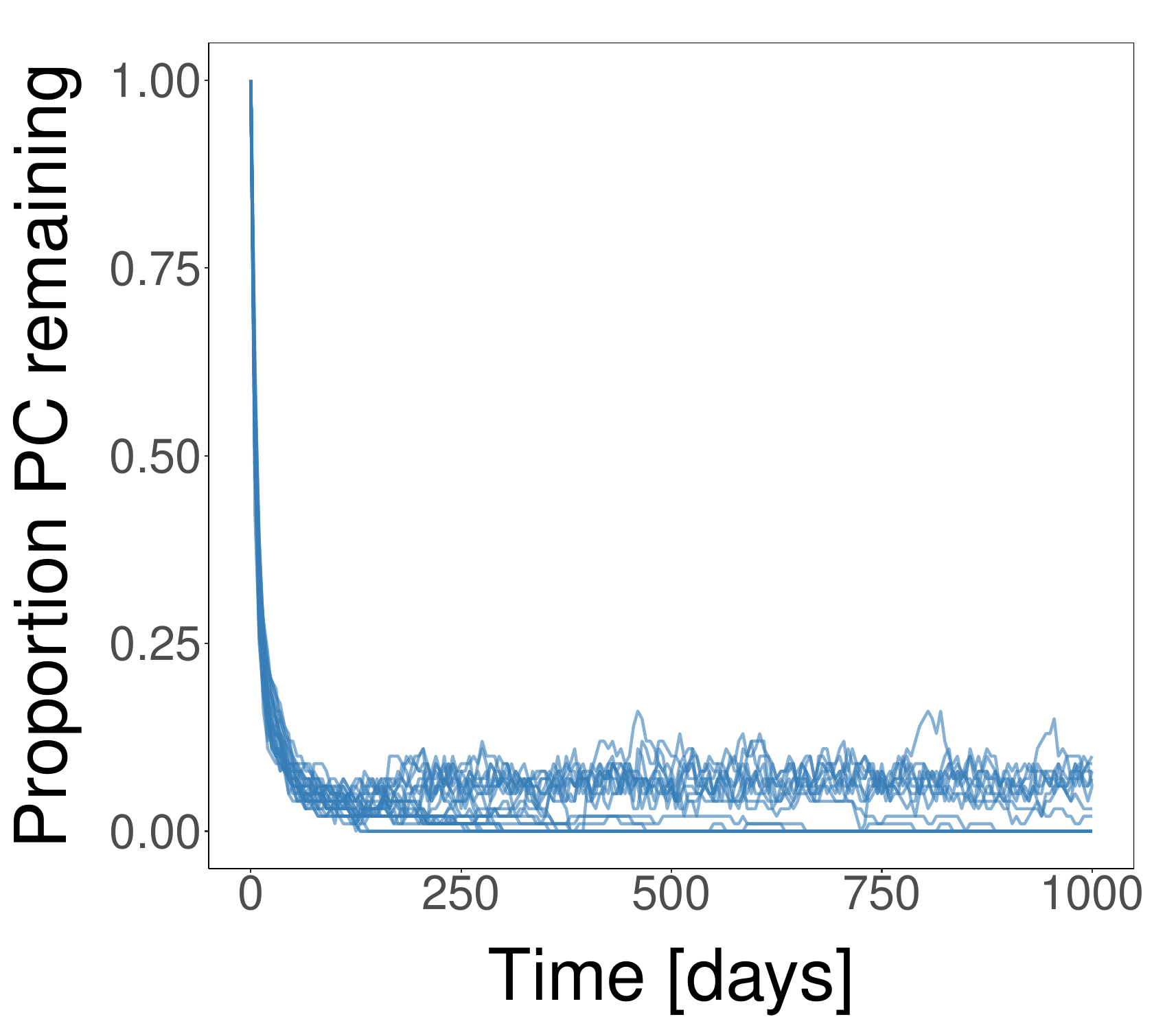}
		\caption{}
		\label{fig:cellloss}
	\end{subfigure}
	\caption{(a) Examples of IFE simulation output. proliferative (Prolif.) cells are shown in red, terminally differentiated (Diff.) cells in blue. The desired proliferative and differentiated structure is shown on the left, and a representative example of proliferative cell loss from the basal layer during simulation on the right. See Supplementary Movie 1 for a movie of the simulation on the right. (b) The proportion of proliferative cells (PC) remaining in the basal layer over time in simulations where the niche is lost. The blue lines are individual instances of the simulation, as there is stochasticity in the system. The black line shows the average number of proliferative cells across the simulation instances.}
	\label{fig:liresults}
\end{figure}

The second mechanistic approach in Panel (C) of Figure~\ref{fig:proliferative_niche} is directed division: the orientation of the direction of the division is regulated.
Directed division is an alternative mechanism that is able to maintain a proliferative cell population of a desired size, limited only by cell volume restrictions, that does not use cell signalling or restrict proliferative cell movement. 
Our results indicate that the method by which division is currently modelled enables easy entry of newly differentiated cells into the niche and therefore causes a loss of proliferative cells from the niche.
Our alternative mechanism, motivated by both these results and the experimentally observed regulation of division orientation, directly opposes the entry of differentiated cells into the basal layer during division by regulating division direction throughout the division process. 

The orientation of division during epidermal development stages has been well studied in mice---divisions parallel to the membrane dominate early development, then a switch to perpendicular occurs later in the development process to aid stratification of the tissue \cite{asare17,kulukian13,lechler05,poulson10,xie17}. 
However less is known about the process in the adult epidermis. 
\citet{pinkus66}, using biopsies of human epidermis, determined 54\% of divisions seen were vertical, 12\% were oblique, and 35\% parallel to the membrane. 
\citet{ipponjima16} imaged division directions in adult mice epidermis and observed a tendency towards parallel or oblique division directions depending on the body region considered. 
Interestingly, they also found that the number of oblique divisions (between 70 and 90 degrees) and the basal cell density (as mentioned previously) was correlated to epidermal thickness. 
Consequently, there is a potential association between oblique divisions and basal cell density. 
Results from \citet{lechler05} found the majority of divisions were perpendicular to the membrane to promote stratification in adult mice, and \citet{sada16} proposed that the majority of divisions of fast cycling stem cells (65\% of stem cells in their model) were coupled to upwards transport of one of the daughter cells.
In contrast, \citet{clayton07} determined that only 3\% of divisions were perpendicular to the membrane in mouse tail. 
Though it is still not clear whether division direction is a significant factor in basal layer homeostasis, these results do imply a regulatory mechanism for division orientation does exist. 

Division direction in biological tissue is determined by the direction the mitotic spindle in the dividing cell, which forms in the final phase of the proliferative cell cycle. 
All proliferative cells undergo a cell cycle consisting of four phases: gap 1 (G$_1$), synthesis (S), gap 2 (G$_2$), and mitosis (M). 
During the M phase the mitotic spindle is formed and the cell elongates. 
The mitotic spindle consists of two poles at either end of the parent cell and determines the division direction of the cell. 
At the end of the M phase, the cell splits at the midpoint between the two poles of the spindle. 

The alignment of the spindle is coupled to cell polarity, which is also known to be important in asymmetric cell division \cite{tervonen11}.
Basal membrane contact and adhesion molecules are important for cell polarity \cite{simons11,tervonen11}, and potentially the membrane provides the trigger that determines the cell's direction of division, as in Drosophila testis and ovaries \cite{simons11}.
In some tissues, misalignment of the spindle can lead to loss of tissue morphology \cite{xie17} and has even been connected to epithelial tumour growth \cite{pease11}.
Given the complexity and many unknowns related to spindle orientation, and how it regulates and maintains its orientation, in this paper we are more interested in its effect: the regulation of the orientation of division.

The regulation of division direction has previously been shown to be useful to include in a mathematical model. 
\citet{gord14} and \citet{du18} found a switch from symmetric division parallel to the membrane to asymmetric division perpendicular to the membrane, combined with polar adhesion, generated correct stratification in their \textit{in silico} subcellular element model of epidermal development. 
Here, polar adhesion means basal and suprabasal regions of cells only form bonds with the basal and suprabasal regions of other cells respectively. 
We propose an alternative approach, appropriate for an overlapping spheres model, of a selected division direction at division and a rotational force to maintain this direction during the M phase of the cell cycle.

In this paper we investigate how the modelling approach influences the potential loss of proliferative cells from the basal layer. 
We then investigate the strategy of the rotational force as a maintenance mechanism for the layer. 
Results show that the inclusion of the rotational force is robust at maintaining a desired proliferative cell population size.
It is likely that the applications for such a force in epithelial tissues is more extensive than just niche maintenance, but for this paper we limit the investigation to its usefulness in maintaining the proliferative basal layer in the epidermis.

\section{The model}

\subsection{Multicellular overlapping spheres model}

We are interested in understanding the phenomenon of loss of the proliferative cell niche in a three dimensional multicellular overlapping spheres model of epithelia. 
Specifically, we look at interfollicular epidermal (IFE) tissue. 
This agent based model approximates cells as spherical agents which are able to overlap. 
In the epidermis we know that keratinocytes are not spherical, they begin in the basal layer as vertical ellipsoids \cite{corcuff93,zhang15} with diameter around 10~$\mu$m, increasing to approximately 15~$\mu$m in the spinosum and 18~$\mu$m in the granulosum \cite{koehler11}. 
In the corneum cells are long and flat with horizontal diameter of 20--40~$\mu$m and a height of less than 0.5~$\mu$m \cite{bouwstra03,kashibuchi02,menon02}.
Given we are interested in the basal and spinous layers, spheres are expected to be a good estimate of the dynamics in these layers where cells are close to spherical. 

Cells interact with each other, and with a membrane boundary, through attractive and repulsive forces. 
The forces we use are based on those used by \citet{li13} and \citet{pathmanathan09}. 
We also introduce a new force to the system: the rotational force during division, which is described at the end of this section.

In the IFE the upward movement of cells occurs due to the division of cells in the basal layer which push differentiated cells upwards through the tissue. 
Cell division is modelled using a stochastic rule-based model \cite{li13,pitt-francis09} where cell cycle duration is randomly chosen from a uniform distribution. 
Of particular interest in this paper is how the mechanics of cellular division are modelled. 
This is described in detail below, along with the interactive forces and the boundary conditions which are experienced by cells.

\subsection{Cell movement through interaction forces}

Cells interact with each other through adhesive forces and repulsive forces. 
Cell movement is then determined by balancing these forces with a drag force. 
Cell inertia is not included as the inertial force term is assumed to be negligible compared to the other forces and cell drag \cite{pathmanathan09}. 
This gives the following equation of motion:

\begin{equation}
	\eta \frac{ \text{d} {\bf r}_i} {\text{d} t} = \sum_{j \in N_i} {\bf F}_{ij} + \sum_{k} {\bf F}_{ik}^{\text{Ext}} \; ,
	\label{eq:forcebalance}
\end{equation}

where ${\bf{r}}_i$ is the cell centre location of cell $i$, $N_i$ is the set of neighbours of cell $i$, ${\bf F}_{ij}$ is the force on cell $i$ due to interacting neighbour cell $j$, ${\bf F}_{ik}^{\text{Ext}}$ are any external forces acting on cell $i$ \cite{pathmanathan09} such as adhesion to the membrane or rotation, and $\eta$ is the drag coefficient for the cell. 
Distance is measured in cell diameters (CD).

We model the force interaction between cells as springs between cell centres, with an optimal, zero-force, spring length of $l_0$. 
We also define ${\bf{\hat{s}}_{ij}}=\lVert {\bf{r}}_j-{\bf{r}}_i \rVert $ the unit vector from cell $j$ to $i$, and  $s_{ij}=|{\bf{r}}_j-{\bf{r}}_i|-l_0$, the deformation of the spring between cells $i$ and $j$. 
The spring length $l_0=1$~CD for mature cells. 
Consequently, the deformation $s_{ij}$ is analogous to the separation, or overlap, between two cell membranes.

The zero-force configuration for two cells is at $s_{ij}=0$~CD. 
When cells are separated they experience forces that pull them together, and when the membranes overlap they experience repulsive forces. 
If cells separate by a distance of $s_{ij}>1.0$~CD, the adhesion molecules are assumed to have ruptured and no force is experienced by the cell pair.
The force functions are given in Equation~\eqref{eq:forces} where the shape parameter $\gamma$ changes the magnitude and distance of the peak force \cite{atwell15,li13,pathmanathan09}:

\begin{align}
\label{eq:forces}
	{\bf F}_{ij} &= \begin{cases}
						-\alpha \left(
  							\left( s_{ij} + c \right) e^{ -\gamma \left( s_{ij} + c \right)^2} - c e^{ -\gamma \left( s_{ij}^2 + c^2 
  							\right)} \right) {\bf{\hat{s}}}_{ij} 
  							& \text{ for } s_{ij} > 0 \; , \\
  						{\bf{0}} & \text{ for } s_{ij} = 0 \; , \\
						k \, {\text{log}}\left(1+s_{ij}\right){\bf{\hat{s}}}_{ij}
							& \text{ for } s_{ij} < 0 \; ,
					\end{cases} \\
	c &= \sqrt{\frac{1}{2\gamma}} \; .
\end{align}

\begin{figure} \centering
	\hfill
	\begin{subfigure}{0.45\linewidth}
		\includegraphics[width=\linewidth]{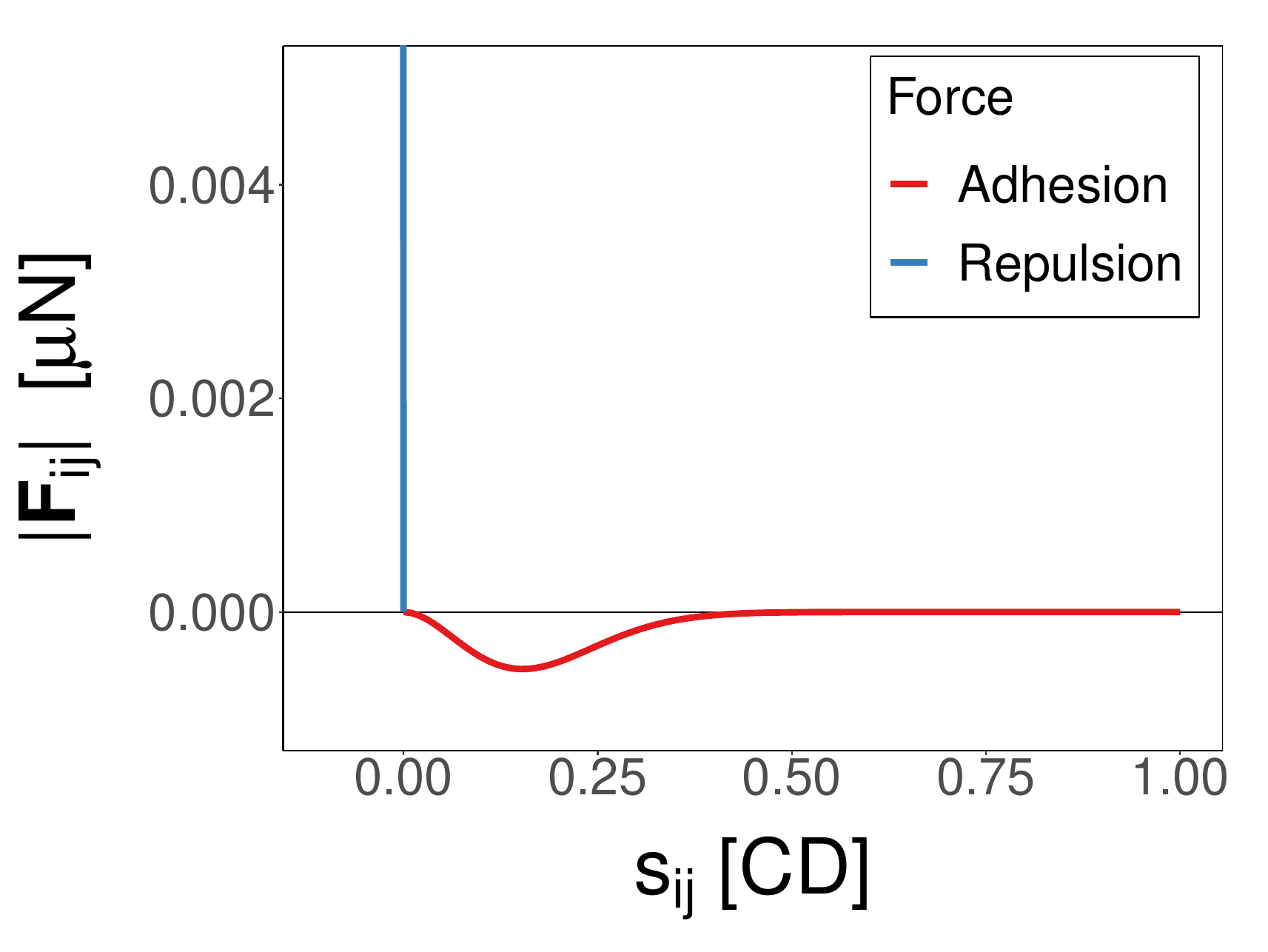}
		\caption{}
		\label{fig:force}
	\end{subfigure}
	\hfill
	\begin{subfigure}{0.45\linewidth}
		\includegraphics[width=\linewidth]{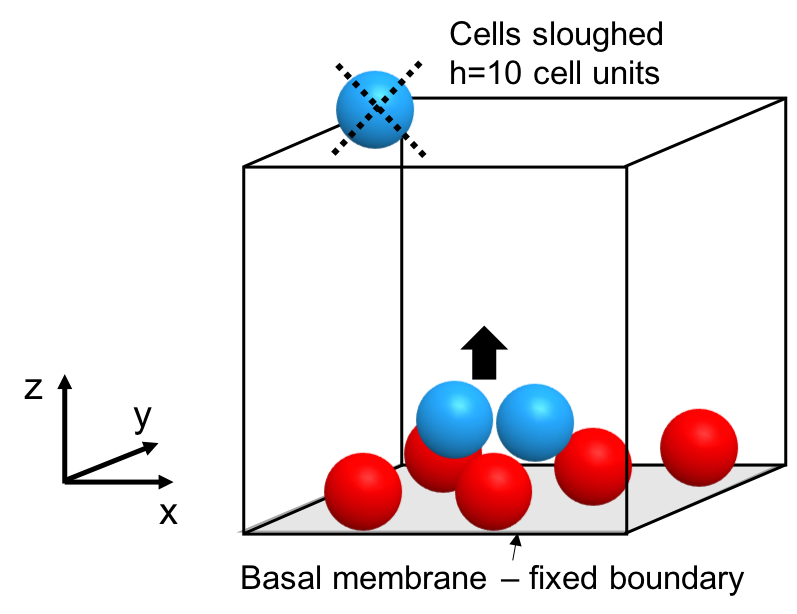}
		\caption{}
		\label{fig:setup}
	\end{subfigure}
	\hfill
	\begin{subfigure}[b]{0.4\linewidth}
		\includegraphics[width=\linewidth]{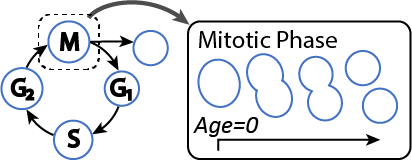}
		\caption{}
		\label{fig:cellcycle}
	\end{subfigure}
	\hfill
	\begin{subfigure}[b]{0.3\linewidth}
		\includegraphics[width=\linewidth]{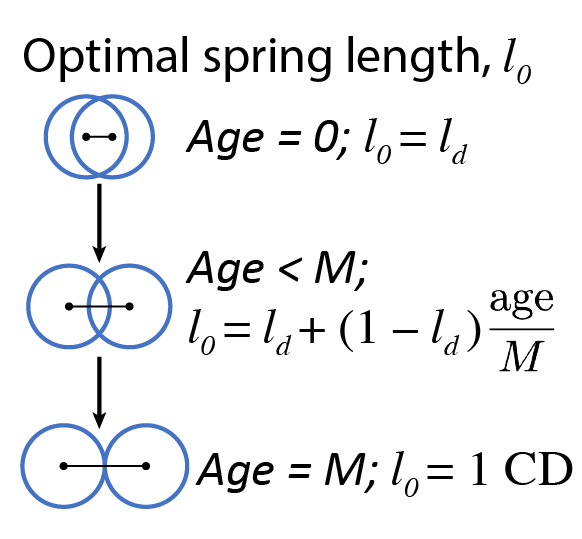}
		\caption{}
		\label{fig:divisionmodel}
	\end{subfigure}
	\hfill
	\begin{subfigure}[b]{0.2\linewidth}
		\includegraphics[width=\linewidth]{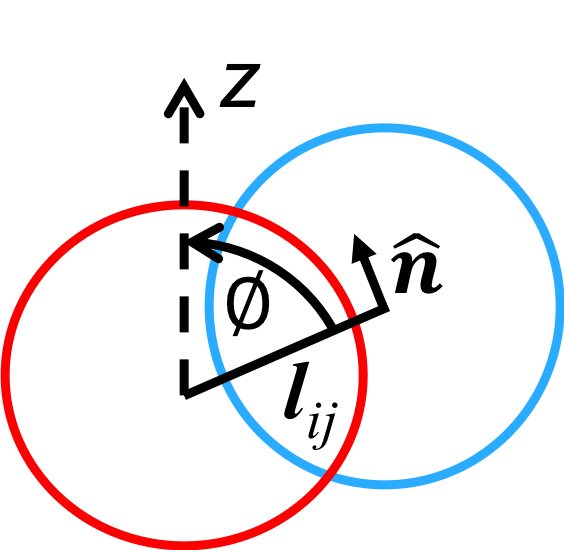}
		\caption{}
		\label{fig:rotationalforce}
	\end{subfigure}
\hfill
	\caption{
		Model diagrams: 
		(a) The force magnitude function for cell overlap ($s_{ij}<0$~CD) and separation ($s_{ij}>0$~CD).
		1~CD$=10~\mu$m.
		(b) Boundary conditions for the domain. Cells move upwards in the z direction and are removed from the top. The horizontal $x$ and $y$ directions have periodic boundaries. 
		(c) The cell cycle phases and detail on the mitotic phase. `$\text{Age}=0$' indicates the starting point of the cell cycle in the model.
		(d) A diagram showing the increasing optimal spring length concept of the model. `Age=0' indicates the point at which the parent splits into two daughter cells with an optimal spring length of $l_d$, which is a parameter we vary within this study. At the end of the M phase, `$\text{Age}=M$', the optimal spring length is the same as between two non-dividing cell (1~CD). Between these two time points the optimal spring length increases linearly with age. 
		(e) The concept of the rotational division force where red is the proliferative cell and blue is the differentiated cell. The z direction is the desired division direction and $l_{ij}$ is the vector between the cell centres. The force is applied along the vector {\bf{$\hat{n}$}}.
	}
\label{fig:model}
\end{figure}

The shape of the force functions are shown in Figure~\ref{fig:force}. 
To avoid overcrowding and un-physical cell densities, we use a logarithmic repulsion ($s_{ij} < 0$) function as opposed to a linear spring function as used in \citet{li13}.

Parameters for the base model are mainly taken from \citet{li13}.
The key parameters for the model are detailed in Table~\ref{tab:allparams}.

\begin{table}
    \centering
    \begin{tabular}{p{0.5\linewidth} l l}
    \hline \hline
        Parameter & Value & Source \\
    \hline
        Cell diameter & 1 CD $= 10~\mu$m & \cite{li13, pathmanathan09} \\
        Cell-cell adhesion coefficient & $\alpha=0.2~\mu$N & \cite{li13} \\
        Adhesion shape parameter & $\gamma=7$ & \cite{li13} \\
        Repulsion spring parameter & $k=150~\mu$N &  \cite{meineke01} \\
        Drag coefficient & $\eta=0.1~\mathrm{\mu N.hr.\mu m^{-1}}$ & \cite{osborne17} \\
        Mitotic phase duration & $M=1$ hr & \cite{alberts15,osborne17} \\
        Base model; division spring length & $l_d=0.001$~CD & \\
        Base model; membrane-proliferative cell adhesion coefficient & $\alpha^*=500~\mu$N & \cite{li13} \\
    \hline
    \end{tabular}
    \caption{Parameters used in the model and their sources.}
    \label{tab:allparams}
\end{table}

\subsection{Adhesion to the basal membrane} \label{sec:bmadhesion}
In addition to forces between cells, we include an adhesive force between cells and the basal membrane.  
Cells will only experience this force if they are `attached' to the membrane. 
Cells attach to the membrane when they come in contact with it. 
They experience the force until they move further than 1.5~CD (cell centre to membrane) above it. 
The adhesive force function between a basal cell and the membrane is the same as that between two cells, but with force coefficient $\alpha^*$ which is stronger than $\alpha$, the force coefficient between cells, as in \citet{li13}. 
The separation distance between the cell and the membrane is the vertical component of the cell's location, $z_i$, which replaces $s_{ij}$ in the adhesive function (Equation~\ref{eq:forces}). 
Unless otherwise specified we use a membrane adhesion coefficient $\alpha^*=500~\mu N$ from \citet{li13}.

The cell membrane is considered an absorbing hard boundary, and the model does not allow any overlap between the cell and the membrane. 
If the force balance on a cell causes it to overlap the membrane, a boundary condition is applied. 
This boundary condition maintains the cell's new horizontal position, but moves the vertical location to zero plus a small perturbation ($\epsilon \sim U(0,0.05)$~CD), which is an approach which has been used previously in other epithelial tissue models that helps prevent over-crowding in the layer \cite{vanleeuwen09}. 
The enforcement of this boundary condition on the cell is what triggers the attachment of the cell, as it has now been in contact with the membrane.

\subsection{Boundary conditions}
The boundary at the base is provided by the absorbing hard boundary membrane as described above. 
The top boundary represents the interface between the epidermis and the outside world. 
As we are interested purely in understanding the activity at the base we implement this in the simplest way possible: by imposing apoptosis on any cell above a certain height. 
This ensures that there is minimal influence from any imposed cell removal on the basal region. 

In this study we use a cube domain with side size 10~CD ($100~\mu m$). 
This is shown in Figure~\ref{fig:setup}. 
We impose periodic boundary conditions on the $x$ and $y$ boundaries to simulate a larger domain. 
Though the simulated tissue height lies at the upper edge of the range of depths of the IFE, it is important to note that, to maintain simplicity, we use spherical cells throughout the height of the tissue and therefore there will be fewer cell layers in the simulation than in real tissue as IFE cells flatten when they migrate towards the tissue surface. 
We do not consider this to pose an issue to the results as the region of interest for this paper is the basal region.

\subsection{Initial conditions}
In order to begin from a homeostatic system, we run an initial fill period to produce a complete tissue. 
The initial proliferative cell count is 100~proliferative cells on a 10~by~10~CD basal plane. 
The system will tend to pack tighter than a 1~CD square for each cell by the end of this fill period, and so the basal layer consists of both proliferative and differentiated cells. 

The fill period uses the same setup as the model described above, with one exception: an extra boundary condition is imposed on proliferative cells to ensure they remain on the membrane. 
This boundary condition only restricts proliferative cell heights, therefore still allowing horizontal movement. 
Once the tissue is filled we remove the height restriction and run the model as described. 
We note that this initial condition is not in an equilibrium state and upon removal of the restriction on proliferative cell locations there is a short initial adjustment period seen as early fluctuations in the results.

\subsection{Modelling mitosis by shifting the clock} \label{sec:modellingmitosis}
When a cell is proliferative it undergoes the cell cycle, as explained in the Introduction, and divides at the end of the M phase. 
In the model we shift this cycle to begin at the start of the M~phase, making $\text{G}_1$ the second phase and so on. 
The biological cycle is shown in Figure~\ref{fig:cellcycle}, with `$\text{Age}=0$' indicating where the model splits the cell into two daughter cells.

Figure~\ref{fig:divisionmodel} shows the cell elongation model after the parent cell has split into two daughter cells and entered the M phase. 
In order to model growth over the M phase, we modify the optimal spring length, $l_0$, between the two daughter cells such that it starts at some length at division $l_d$ and increases linearly in time over an hour until $l_0=1$~CD at the end of the M phase \cite{almet18,drasdo03,meineke01,pitt-francis09}. 
The smaller the value of $l_d$ the more physically realistic the model as it more accurately reflects the elongation of the cell and ensures a smooth transition for adding volume to the system. 
The values used for the plots in Figure~\ref{fig:liresults} is $l_d=0.001$~CD, which is the smallest spring length used in this study and is essentially the same as beginning with a single cell, which then stretches into an oblong shape with a pinched centre before becoming two separate cells.
The larger value of $l_d$ (0.1~CD) means at the start of the mitotic phase a cell is added to the system with a small displacement, adding a small discontinuity to the system. 
We note that the dynamics of the system would be unchanged if instead we grew the cell into an oblong shape and then divided it into two cells at the end of the M phase. 
We use this implementation as it is easily implemented in an overlapping spheres model.
Another approach is to grow the cell in a spherical shape, and then split into two, as in \citet{galle05}, however we consider growth in an oblong direction to be more biologically realistic \cite{alberts15,ipponjima16}.

Though the spatial location of the proliferative cells in the epidermis is known, the cell lineage is not completely understood. 
For example, it is not known whether one or more different types of proliferative cells are active in replenishment of the tissue, and additionally whether cells divide symmetrically to produce cells of the same type or asymmetrically to produces cells of different types \cite{clayton07,kaur11,li13}. 
Originally it was thought that stem cells divide asymmetrically to produce transit amplifying cells which further differentiated after a series of symmetric divisions \cite{potten87}.
Recent studies support population asymmetry models in which cells divide both symmetrically and asymmetrically, either within the same pool of stem or progenitor cells \cite{clayton07,mascre12} or from two distinct pools of progenitor cells: one pool dividing symmetrically and the other asymmetrically \cite{sada16}.
For simplicity we only use asymmetric division in this paper---a proliferative cell produces one daughter proliferative cell and one non-proliferative, or terminally differentiated, daughter. 
This model is the equivalent of any asymmetrically dividing population, either with or without intermediate transit amplifying or progenitor cells, as long as the number of proliferative cells producing differentiated cells remains constant.
The only difference would be a scaling of the division rate.
This approach does not incorporate dynamics for a symmetrically dividing population, as this is much more complicated and will be future work for the model, however in all studies the majority of cells were found to divide asymmetrically (65\%-84\%) \cite{clayton07,mascre12,sada16} and so this is the model used in the paper.

At the point of creation of the two new cells the cells are placed into the desired vertical alignment with the differentiated daughter above the proliferative cell daughter. 
We note even this provides a preliminary model for directed division, though it does not guarantee the direction is maintained during the mitotic phase.
In the base model there are no additional forces or restrictions applied to the cells after they enter the system.

\subsection{The rotational force during cell division}
We propose to apply an additional rotational force to the differentiated daughter cell during the M phase of the division model as a mechanism to help maintain the proliferative cell population. 
As discussed in the Introduction, there is evidence that supports the regulation of division direction in the epidermis.
We introduce this force as a phenomenological way to capture the cells ability to orientate its division, replicating the behaviour that would be seen if directed division occurred in the epidermis.
This correctional force opposes the mechanism by which we expect cells are lost from the niche in current models. 

As the actual mechanism for the division orientation is highly complex and not completely understood, in this paper we implement a simple phenomenological model to emulate the effects of the mechanism at a cellular, rather than a subcellular, level.
It is important to note that, with this model, we are aiming to reproduce the dynamics of the system if directed division occurs in the epidermis and are not explicitly modelling any observed force due to the spindle.
For this model we made the assumption that division orientation is an asymmetric process, with the proliferative daughter remaining in place on the membrane, and hence only apply the force to the differentiated daughter.
Arguments could equally be made for applying the force to both of the daughter cells, however such a model would produce the same qualitative results as the majority of any force applied to the proliferative daughter would be neutralised by the boundary condition due to the membrane and only result in small perturbations in position on the membrane. 

We base our model on a torsional spring force that is independent of spring length, to mirror the simplest interaction force, the linear spring force, used in agent based models \cite{pathmanathan09}.
This force is in addition to the vertical alignment of the daughter cells at the point of division described above, and the daughter cells still continue to experience the usual repulsion and attraction forces.
Given we use an asymmetric division lineage, the proliferative cell should remain attached to the membrane and the differentiated cell should enter the suprabasal layers.
Figure~\ref{fig:rotationalforce} illustrates the concept of the force---rotating the differentiated daughter cell towards the upwards vertical position. 
The force equation is given in Equation~\eqref{eq:torsionalspring} where $k_{\phi}$ is the torsional spring constant, $\phi$ is the angle (in radians) between the division vector and vertical, and ${\bf{\hat{n}}}$ is the unit normal to the division vector: 

\begin{equation}
	\label{eq:torsionalspring}
	{\bf{F}}_{i}^{\text{Rot}}= -k_{\phi} \phi {\bf{\hat{n}}} \;.
\end{equation}

We have explicitly defined a function of $\phi$ to account for the case when the differentiated cell drops below the proliferative cell. 
The use of a function that is independent of the length of $\mathbf{s}_{ij}$ is because a 2--3~orders of magnitude variation of $\mathbf{s}_{ij}$ over the mitotic time period results in 2--3~orders of magnitude variation in the determined force and hence issues with numerical stability.
It would be feasible to use an alternative function of $\phi$ (explicitly or otherwise) to the linear function shown here that produces similar behaviours, however for any reasonable increasing function of $\phi$ qualitatively similar results are expected. 

The force is applied, to the differentiated daughter only, in a direction normal to the separation vector of the two cells and towards the desired vertical direction in the plane of the separation vector and the desired vertical direction at each time step. 
This is defined mathematically as:

\begin{equation}
	{\bf{\hat{n}}} = \frac{\mathbf{s}_{ij} \times ( \mathbf{k} \times \mathbf{s}_{ij} )}{\norm{\mathbf{s}_{ij} \times ( \mathbf{k} \times \mathbf{s}_{ij} )}} \mathrm{,}
\end{equation}

where $\mathbf{s}_{ij}$ is the unit vector between daughter cells $i$ and $j$, and $\mathbf{k}$ is the vertical unit vector, representing the desired division direction, noting that ${\mathbf{\hat{n}}\times\mathbf{s}_{ij}=0}$. 
The angle $\phi$ is consequently bounded by $[0,\pi]$, with $\pi$ being the highly unlikely case that the differentiated cell is pushed vertically below the proliferative cell. 

\subsection{Implementation}
\sloppy
The model is developed using the open source Chaste software libraries \cite{mirams13,osborne17,Cooper2020}. 
Chaste is a C++ library used to run cardiac and multicellular tissue simulations. 
The core Chaste code can be accessed from \url{https://chaste.cs.ox.ac.uk/trac/wiki}. 
Additional code to reproduce the results in this paper can be found at \url{https://github.com/clairemiller/2021_MaintainingProliferativeCellNicheEpithelia.git}.

\subsection{Steady state population estimation}

In order to compare the effectiveness of the different parameter combinations at maintaining the niche, it is necessary to determine a comparison metric. 
Motivated by the decay curves observed (see Figure~\ref{fig:cellloss}) we assume the proliferative cell loss from the niche approximately follows an exponential form: cells detach from the basal membrane at some rate $\lambda$, but with a steady state population $\beta$ of the proliferative cells remaining attached to the membrane.
This is described by the following function:

\begin{equation}
	p(t) = \omega e^{-\lambda t} + \beta \; ,
	\label{eq:lossfit}
\end{equation}

where $p(t)$ is the population of proliferative cells attached to the membrane at time $t$, and $\omega$ is the lost population. 
We define attached cells, and hence cells in the basal layer, using the definition in Section~\ref{sec:bmadhesion}: the cell has made contact with the basal layer and has not separated more than 1.5~CD from its contact point.
Physically, $\beta$ represents the number of proliferative cells that remain in the basal layer long term, and is consequently an estimate for the maximum density of proliferative cells the system can maintain in the niche.
Given we simulate a horizontal domain of size $10^4~\mu m^2$, the expected maintained proliferative cell density in the basal layer is given by $\beta$~cells/$10^4~\mu m^2$. 
The fit of this model to the simulation data depends on the parameter values and hence $\beta$ is a useful comparison metric for the different parameter combinations.

At very high steady state populations, when the mean proliferative cell loss was less than 6\%, the regression algorithm could not find an appropriate fit to the data. 
This is due to the assumption of an initial transient, given by $\omega e^{-\lambda t}$, in Equation~\ref{eq:lossfit}.
When the starting population is very close to the steady state population, the algorithm fails to find a transient to fit.
Investigating these results by calculating the fit for $\beta$ with a designated $\lambda$ shows little variation in the calculated $\beta$, implying the data is relatively insensitive to $\lambda$ over the time periods used.
Consequently, in the cases where the mean loss was less than 6\%, we instead set $\beta$ as the mean proliferative cell count, at the final time point, of the 25~realisations run.

\section{Results}

We include stochasticity in both the cell cycle time and the membrane boundary, as detailed above. 
Consequently, for each parameter set we run the setup with 25~seeds. 
Results show either both the individual realisations and the mean, or just the mean, as stated in the relevant caption.

\subsection{Loss of the niche affects system dynamics of the tissue}
Using the model without the rotational force, which we will refer to as the base model, we see in Figure~\ref{fig:liresults} that the system is unable to maintain a proliferative cell density in the basal layer greater than 16~cells/$10^4~\mu m^2$.
Additionally, as mentioned previously, 56\% of the simulations have no proliferative cells from the niche at day~1,000. 
The number of proliferative cells in the basal layer is determined by the number of proliferative cells which are experiencing the attachment force with the membrane.
Consequently it is possible for a cell to detach and then re-attach later, which is why the lines do not always monotonically decrease in Figure~\ref{fig:liresults}.

\begin{figure} \centering
	\begin{subfigure}{0.5\linewidth}
		\includegraphics[width=\linewidth]{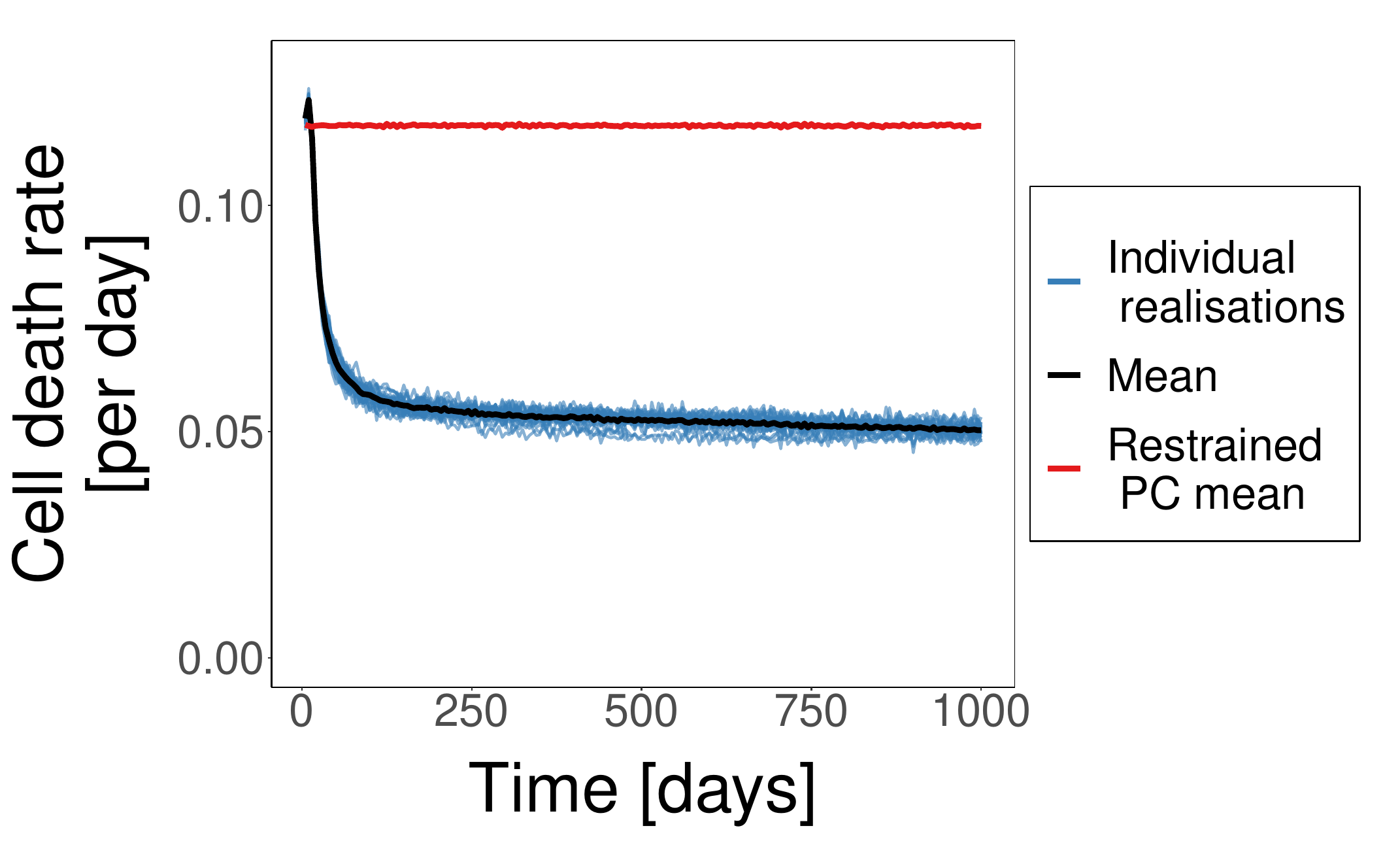}
		\caption{}
		\label{fig:celldeath}
	\end{subfigure}
	\begin{subfigure}{\linewidth}
		\includegraphics[width=\linewidth]{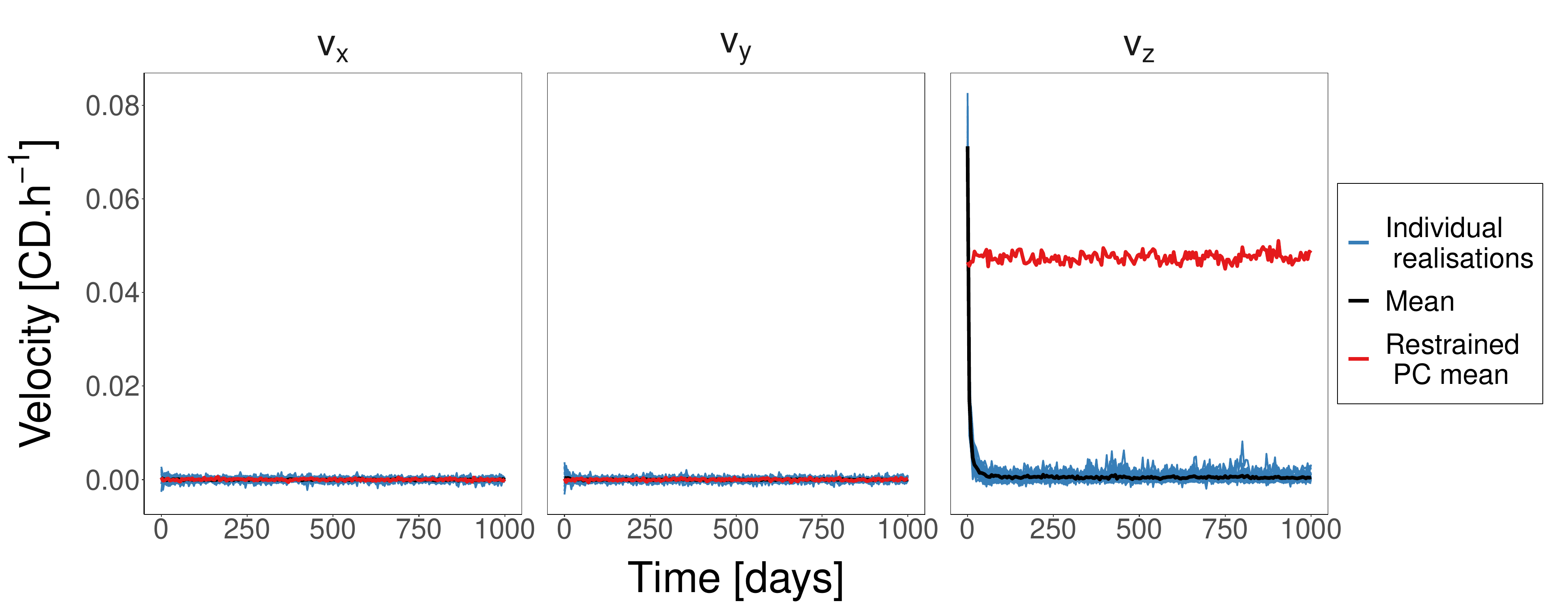}
		\caption{}	
		\label{fig:velocities}
	\end{subfigure}
	\caption{The effect of the loss of the proliferative cell niche on system dynamics: (a) cell deaths from the top of the tissue, as a proportion of total cell count, and (b) the cell velocities, in units of cell diameters (CD) per hour, in each direction. Blue lines show the individual simulation instances, and the black line is the average. The red line shows the comparative results for a simulation with vertically restrained proliferative cells (PC).}
	\label{fig:systemdynamics}
\end{figure}

In addition to disruptions to tissue structure, proliferative cell detachment affects system dynamics such as cell deaths at the top of the tissue and cell velocities. 
These results are shown in Figure~\ref{fig:systemdynamics} for the base model compared to a model in which proliferative cells are vertically restrained to remain on the basal membrane, which represents the expected homeostatic system. 
The initial overshoot seen in the death rates and velocities is because the tissue at the end of the fill period is not in exact equilibrium, and so there is a short initial settling phase when the proliferative cell restriction is released. 
As time increases, and proliferative cells are lost, the vertical velocity and cell death rate decreases. 
This would be expected as the decrease in proliferative cells causes a decrease in divisions. 
Consequently, less upwards force is exerted on cells, decreasing velocity and slowing the rate at which cells reach the top boundary.  
Slower upward velocities and lower death rates can dramatically change simulation dynamics, such as cell turnover times.

\subsection{Only a very high level of membrane adhesion can maintain a niche in the base model}
In order to fully investigate the problem we try different parameter combinations for the division and membrane adhesion parameters in the model. 
The different parameter ranges investigated are shown in Table~\ref{tab:sweepparameters}. 
Doubling the length of the M phase, from 1 to 2 hours, was also tried, however the difference in results was minimal and is not included here for brevity.
 
\begin{table} \centering
	\begin{tabular}{c p{0.45\textwidth} r}
		\hline\hline
		Parameter & Description & Value range \\
		\hline
		$\frac{\alpha^*}{500}$ & Basal membrane adhesion, as a multiple of the base model membrane adhesion & 0--3 \\
		$l_d$ & Starting spring length at division [CD] & $10^{-3}$--$10^{-1}$ \\
		\hline\hline
	\end{tabular}
	\caption{Sweep parameters.}
\label{tab:sweepparameters}
\end{table}

\begin{figure}
\begin{subfigure}{0.49\linewidth}
	\includegraphics[width=\linewidth]{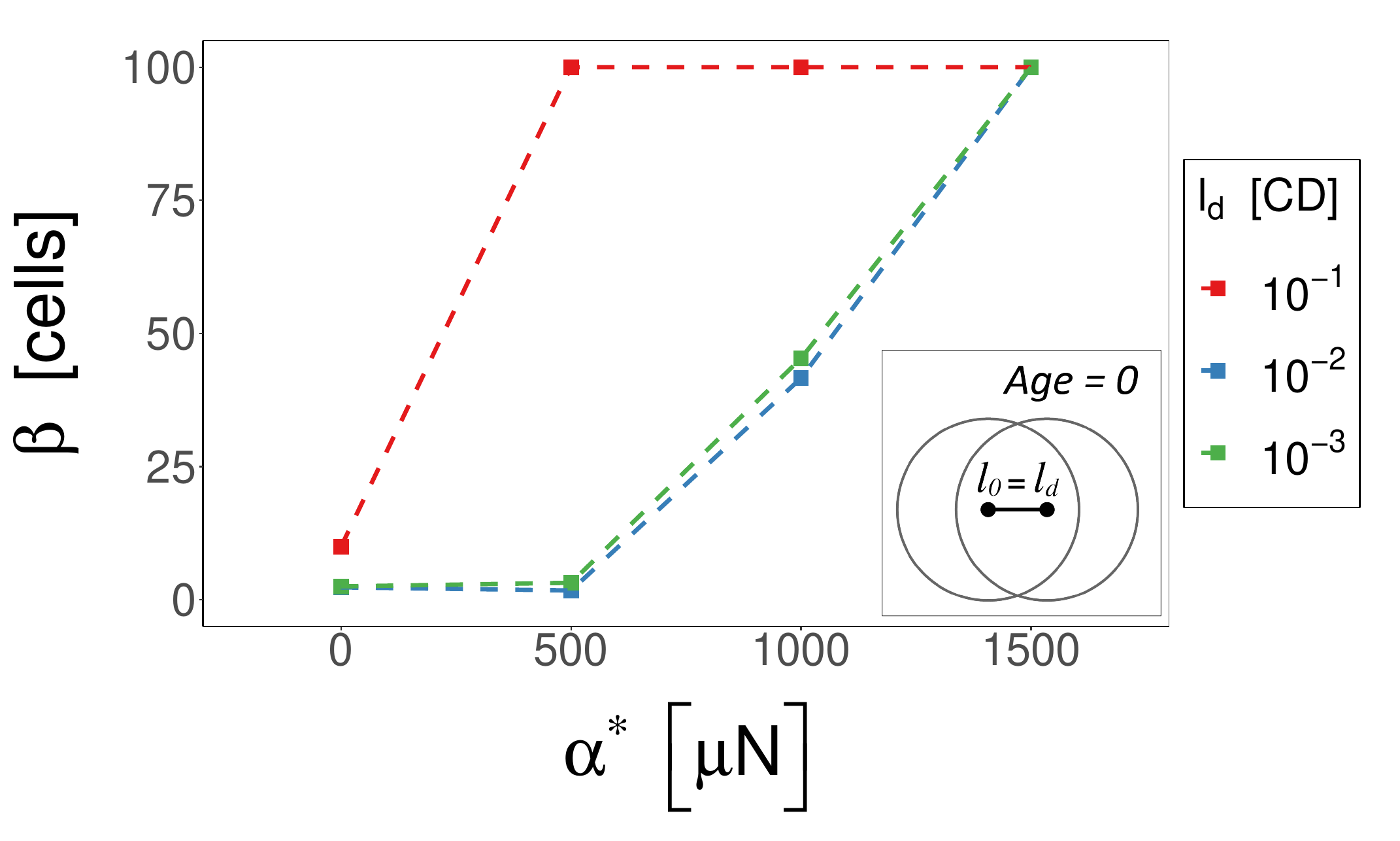}
	\caption{}
	\label{fig:baseremains}
\end{subfigure}
	\hfill
\begin{subfigure}{0.49\linewidth}
	\includegraphics[width=\linewidth]{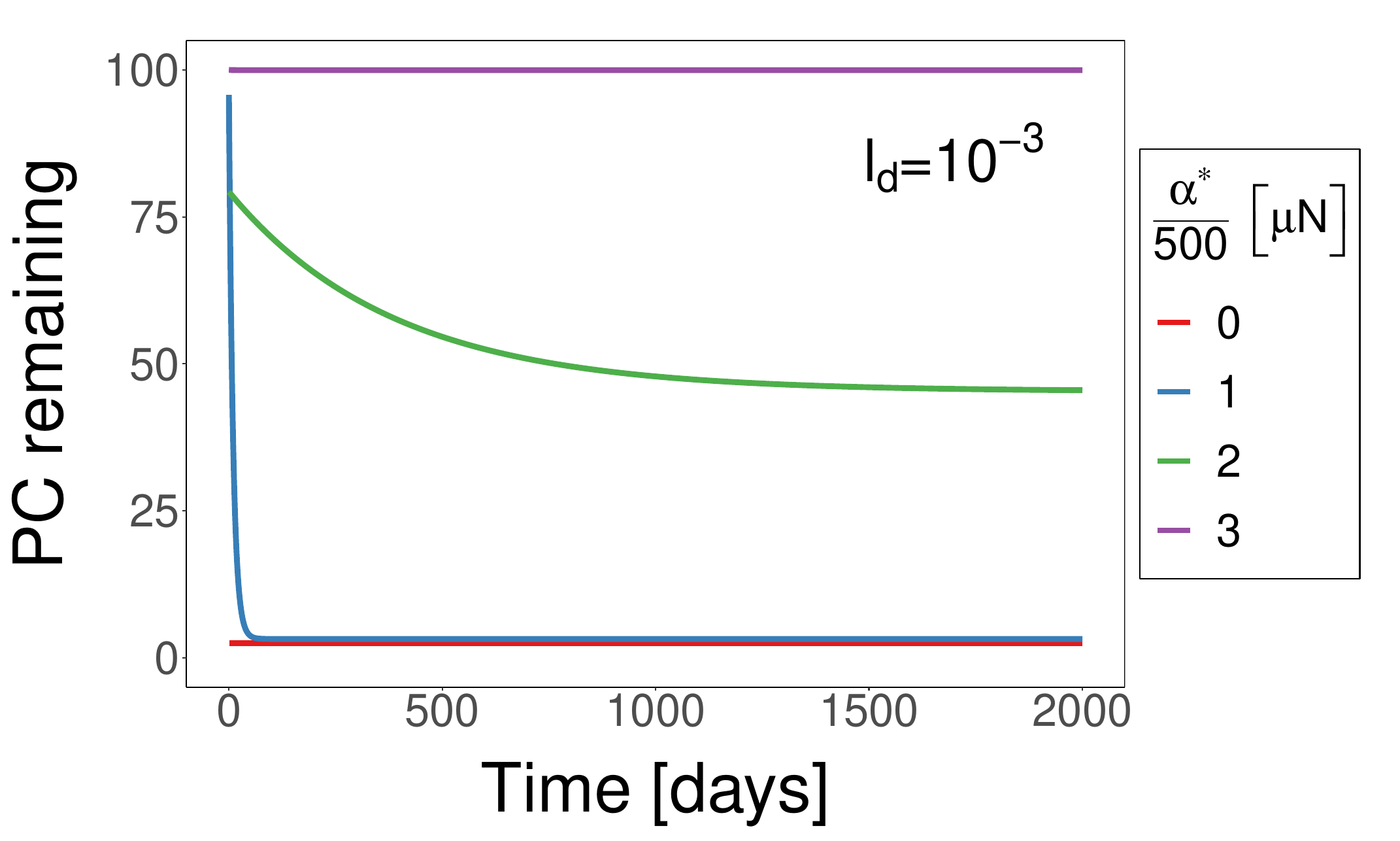}
	\caption{}
	\label{fig:baseloss}
\end{subfigure}
	\caption{
	(a) The value of $\beta$, the remaining proliferative cell population in the niche, for each parameter combination. The unit of $l_d$ is cell diameters (CD). The inset reiterates the model definition of $l_d$, as explained in Figure~\ref{fig:divisionmodel}.
	(b) Examples of the loss curves for the different levels of adhesion to the basal membrane at the lowest division spring length considered; $l_d=10^{-3}$~CD. PC: Proliferative cells.}
	\label{fig:parametersweep}
\end{figure}

The values of the estimated steady state population, $\beta$, for the different parameter combinations are shown in Figure~\ref{fig:baseremains} alongside examples of fits in Figure~\ref{fig:baseloss}. 
The plot in Figure~\ref{fig:baseremains} shows that increasing the adhesion to the basal membrane significantly increases the remaining proliferative cell population. 
It can also be seen that the remaining proliferative cell population decreases as the spring length at division ($l_d$) decreases. 
As stated in the methods, the division spring length parameter determines the translation length at which the new cell is added to the system.
The smaller this parameter, the smoother and more realistic the growth and then division of the cell is modelled.
Figure~\ref{fig:baseremains} shows that at small spring lengths it is necessary to increase the adhesion to three times the value of \citet{li13} to maintain the desired population size in this paper. 
Consequently, we see that we are unable to maintain desired proliferative cell densities at low division spring lengths without high levels of adhesion to the membrane.

\subsection{Proliferative cell loss is due to neighbouring cell interactions during division}
\begin{figure}
\begin{subfigure}{0.4\linewidth}
		\includegraphics[width=\linewidth]{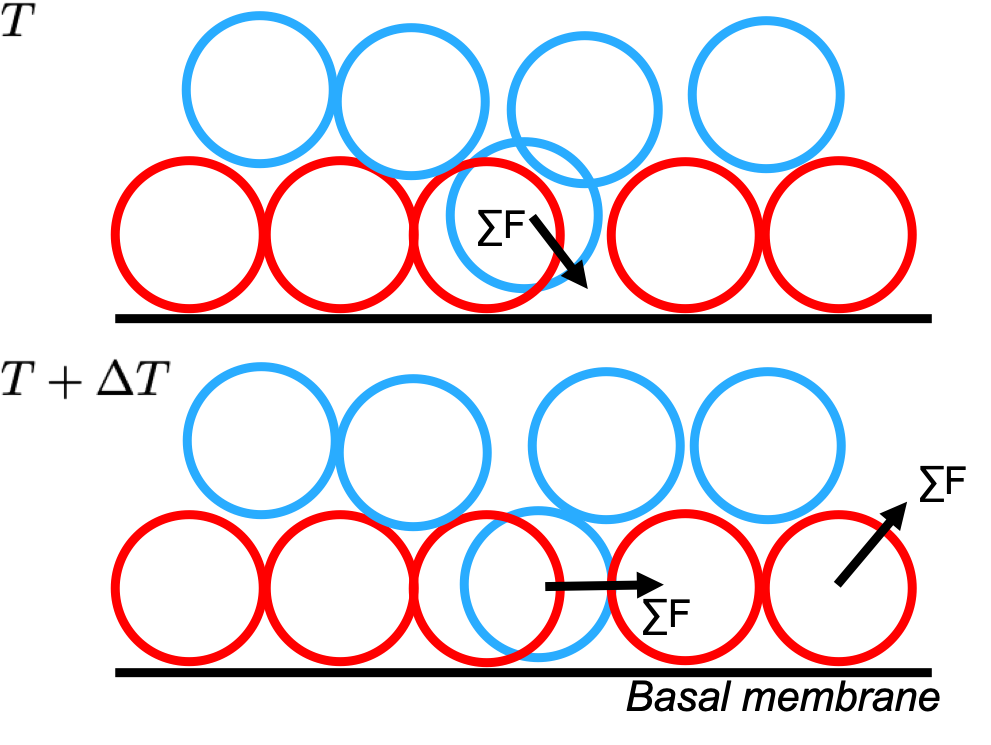}
		\caption{}
		\label{fig:diagramcellloss}
\end{subfigure}
\hfill
\begin{subfigure}{0.55\linewidth}
		\includegraphics[width=\linewidth]{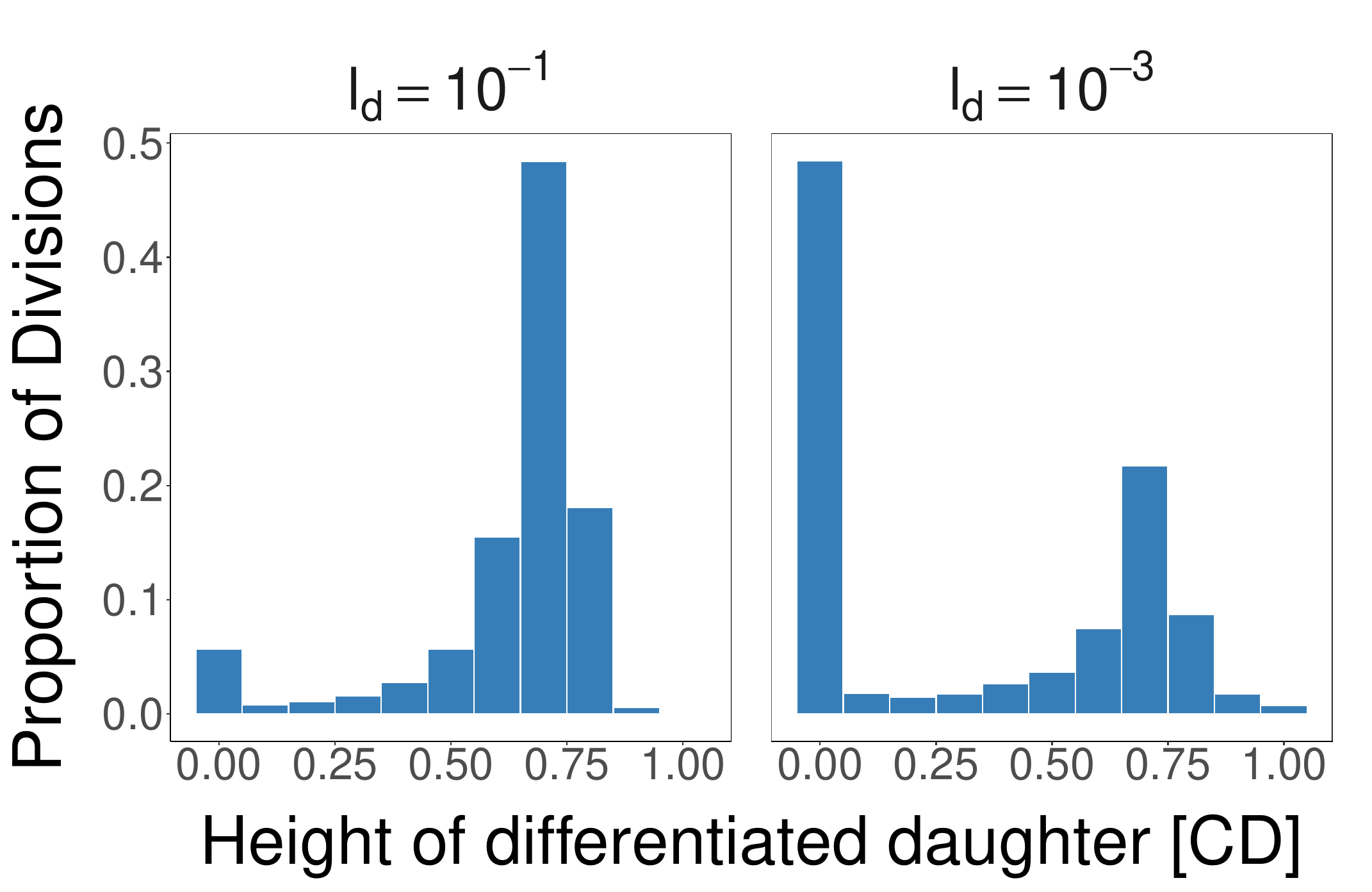}
		\caption{}
		\label{fig:heightatdivision}
\end{subfigure}

	\caption{(a) The proposed method by which differentiated cells enter the basal layer and push proliferative cells out. proliferative cells are shown in red, terminally differentiated in blue. The top image shows the system at time $T$, immediately after the division has occurred. The lower image shows the system at time $T+\Delta T$, where $\Delta T$ is less than the M phase length and so the cells are still within the M phase. (b) A histogram of the height of the differentiated cell's centre at the end of M phase for the largest (left) and smallest (right) spring lengths at division. The unit of $l_d$ and the height is cell diameters (CD).}
\end{figure}

We wish to understand the underlying mechanism for cell loss from the basal layer. 
We propose that it is related to the low displacement required to rotate the daughter cell into the basal layer at the smaller spring lengths during the M phase of the division model. 
This idea is reinforced by the lower remaining proliferative cell population ($\beta$) seen as the division spring length ($l_d$) is decreased in Figure~\ref{fig:baseloss}. 
Decreasing the division spring length decreases the displacement required to move the daughter into the basal layer.
Once the daughter is in the basal layer, the layer becomes overcrowded and a cell must be pushed out. 
If the suprabasal area above a proliferative cell is at a lower density than that above the newly basal differentiated cell, the proliferative cell will be pushed out. 
This is because, in the model, the attractive forces between the proliferative cell and the membrane are lower than the repulsive forces between the suprabasal cells and the differentiated daughter. 
This is shown diagrammatically in Figure~\ref{fig:diagramcellloss}. 

In order to investigate the validity of this proposed mechanism we can determine the number of differentiated daughter cells in the basal layer at the end of their M phase. 
The end of the M phase is the point in time at which the spring length between the two daughters reaches one cell diameter.
This is also when the two daughter cells become two separate cells and are no longer considered a dividing pair. 
Figure~\ref{fig:heightatdivision} shows a histogram of the average heights over the first ten days of simulated time using the base model for the smallest ($10^{-3}$) and largest ($10^{-1}$) values of $l_d$ considered, with $\alpha^*=500~\mu$N. 
If the differentiated cells are being pushed into the basal layer as proposed we would expect to see large numbers of cells at, or close to, zero. 
As can be seen on the right in Figure~\ref{fig:heightatdivision}, there are more differentiated daughter cells in the basal layer at the smaller division spring length, $l_d=10^{-3}$. 
These results support the idea that the smaller spring lengths enable the rotation of differentiated daughter cells into the basal layer, hence pushing proliferative cells out. 

It is interesting to note that a mode in each of the histograms is seen around 0.7~CD. 
We would expect the cell packing to form a triangular pyramid style of packing, which would place the second layer of cells at roughly 0.8~CD, slightly higher than seen in this plot.
The lower height is likely due to the compression both in the system overall, and particularly around a division event as the daughter cells press into the surrounding cells during the division.

\subsection{A rotational force improves maintenance of the niche}
\begin{figure} \centering
	\begin{subfigure}{0.8\linewidth}
		\includegraphics[width=\linewidth]{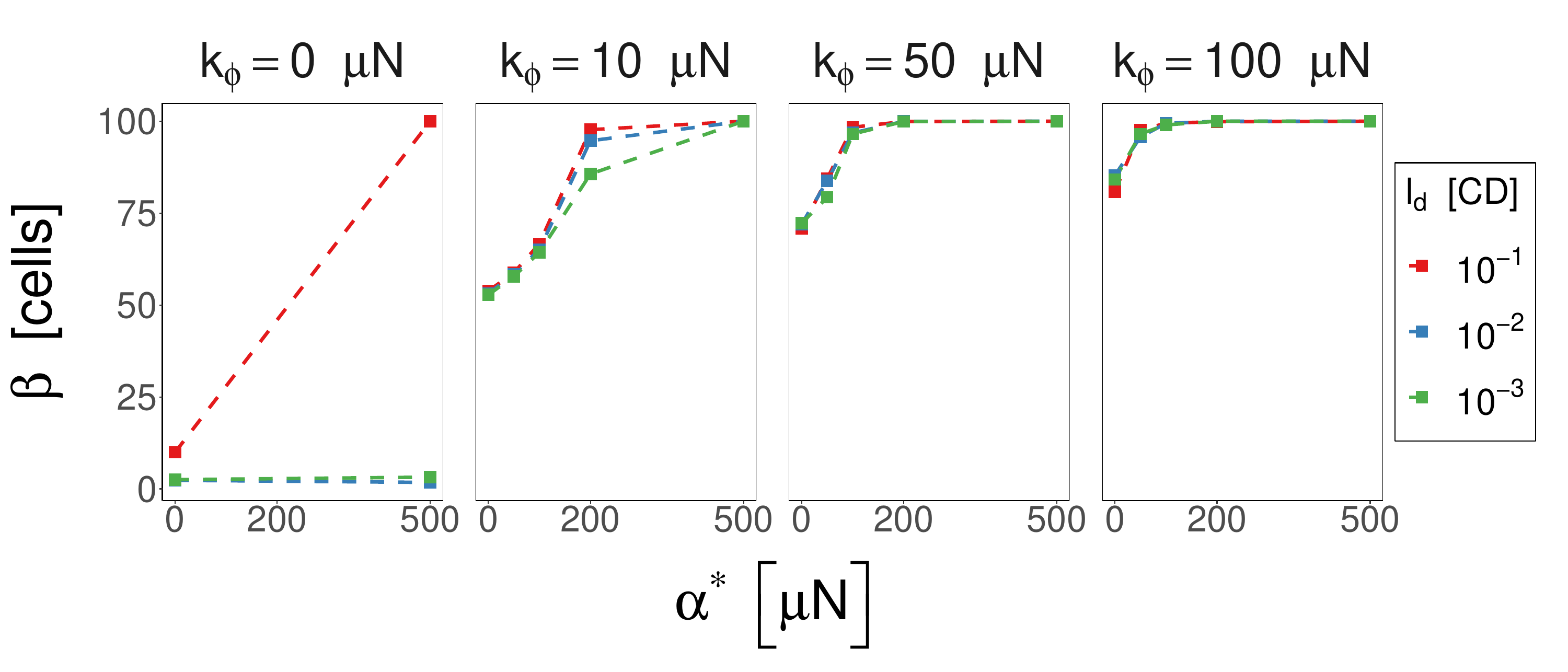}
		\caption{}	
		\label{fig:rotremains}
	\end{subfigure}
	\begin{subfigure}{0.55\linewidth}
		\includegraphics[width=\linewidth]{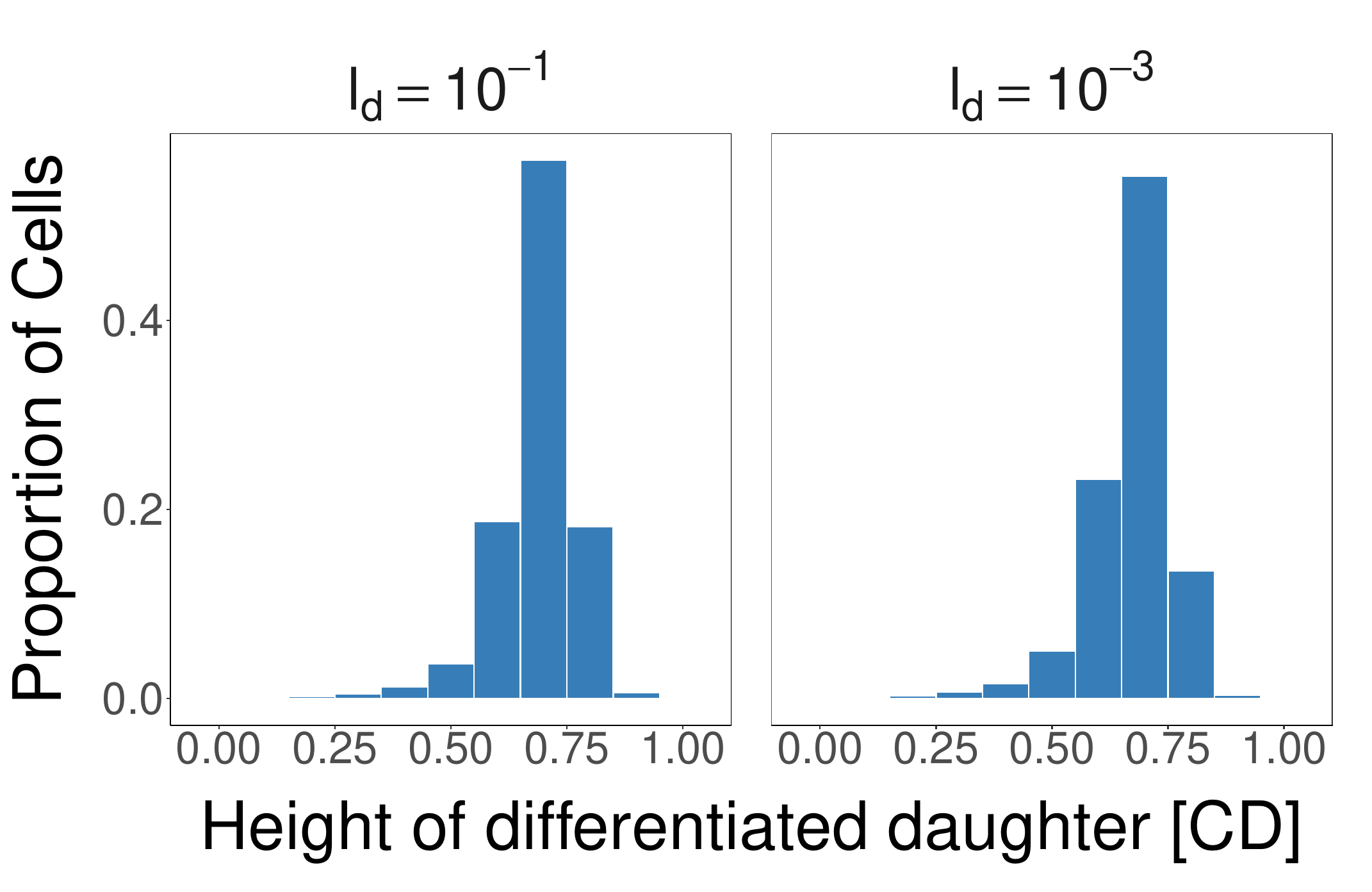}
		\caption{}	
		\label{fig:rotcellheights}
	\end{subfigure}
	\caption{(a) The results for the rotational force during the mitotic phase of the cell cycle model. Adhesion levels shown are $\alpha^*=$ 0, 50, 100, 200, and 500~$\mu$m. The leftmost panel ($k_\phi=0~[\mu N]$) are the base model results (no rotational force) for comparison.
	(b) The height of the differentiated cell's centre at the end of the M phase of division for the largest and smallest division spring lengths. The rotational spring constant used for these plots was the smallest tested, $k_{\phi}=10 \; \left[ \mu N \right]$, and the membrane adhesion was the same as the base model, $\alpha^* = 500 \; \left[ \mu N \right]$. The unit of height is cell diameters (CD).
	}
\end{figure}

Motivated by the observed regulation of division orientation we impose a rotational force on the differentiated daughter cell during the M phase, which directly opposes the proposed mechanism of cell loss described above. 
In Figure~\ref{fig:rotremains} we plot the estimated steady state population, $\beta$, for simulations using the rotational force.
We present results for basal adhesion ($\alpha^*$) levels between zero and the adhesion level used by \citet{li13} ($500~\mu N$).

Figure~\ref{fig:rotremains} shows that including the rotational force increases the population of proliferative cells that remain attached to the membrane ($\beta$) for all spring lengths ($l_d$), helping to maintain the desired proliferative cell population size. 
With the base level of adhesion to the membrane ($\alpha^*=500~\mu$N), any of the rotational spring constants investigated ($k_\phi = 10,~50,~100~\mu$N) maintain the whole proliferative cell population. 
With no adhesion and a torsional spring constant of $k_\phi = 50~\mu N$ or higher, the remaining population is maintained above 70~cells/$10^2~\mu m^2$. 
Supplementary Figure~1 shows a plot of the simulation loss curves for each setup considered using the rotational force. 

It can be seen in Figure~\ref{fig:rotcellheights}, compared to Figure~\ref{fig:heightatdivision}, that the rotational force has significantly reduced the presence of differentiated daughters in the basal layer, directly opposing our proposed mechanism for why the proliferative cell loss occurs. 
It is also important to note that the mode of the histogram remains around $z=0.7$~CD, we have only removed the  peak at $z=0$~CD. 
Consequently, the addition of the rotational force has improved the maintenance of the basal layer over the time periods of interest, even at very small division spring lengths where the division model is most representative of cell division and the base model requires high adhesion levels for maintenance.

\begin{figure} \centering
	\begin{subfigure}{0.45\linewidth}
		\includegraphics[width=\linewidth]{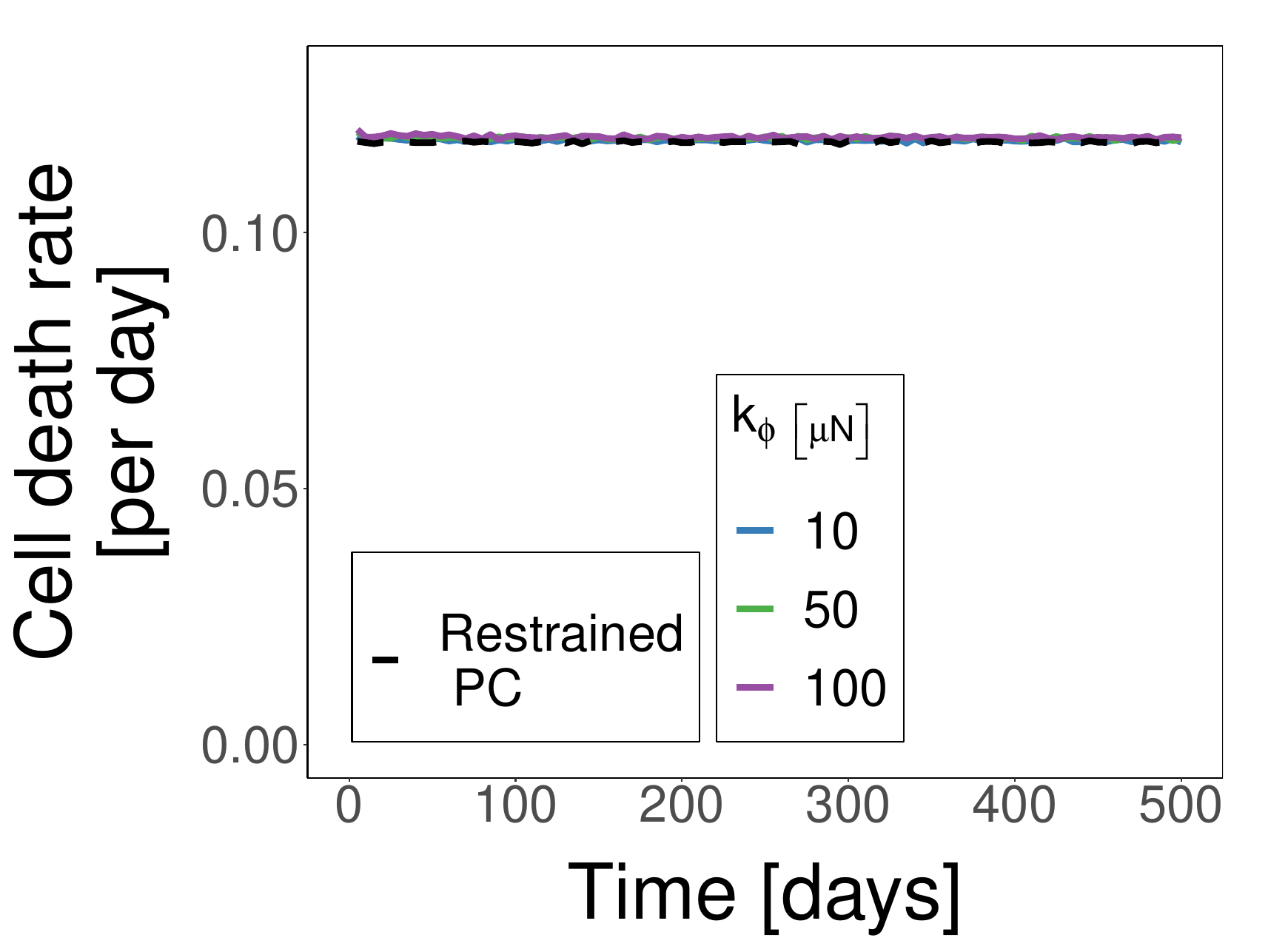}
		\caption{}
		\label{fig:rotcelldeaths}
	\end{subfigure}
	\begin{subfigure}{0.75\linewidth}
		\includegraphics[width=\linewidth]{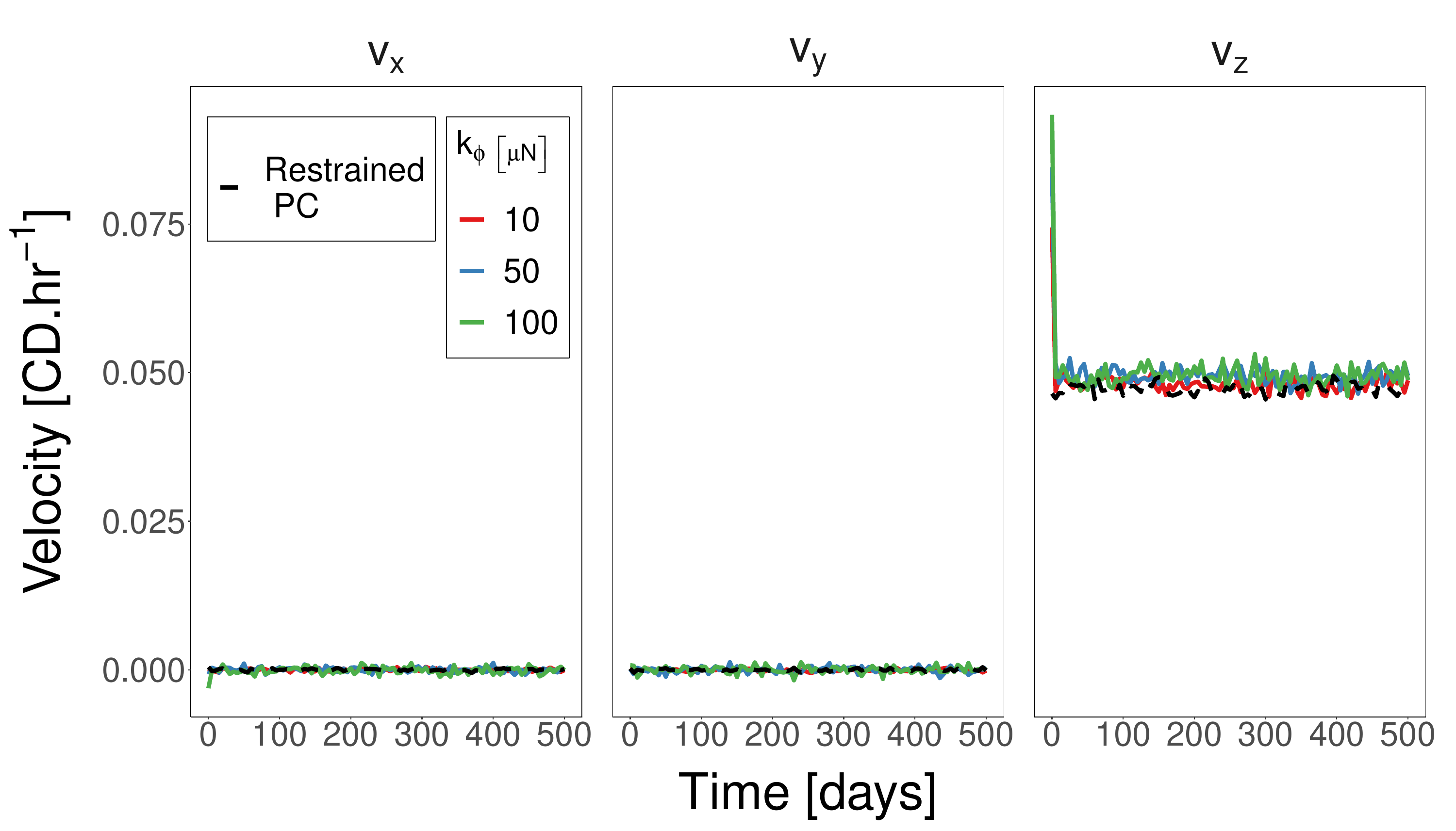}
		\caption{}
		\label{fig:rotcellvelocities}
	\end{subfigure}
	\caption{The new system dynamics with the rotational force included: 
	(a) The cell death rate, as a proportion of the total cell count, and 
	(b) velocities for the different rotational forces when adhesion to the membrane is $\alpha^*=500~\mu N$ and starting spring length at division is $l_d=10^{-3}$~CD. The dashed black line shows the comparative results for a simulation with vertically restrained proliferative cells (PC). The unit of velocity is cell diameters (CD) per hour.
	}
\end{figure}

The cell deaths and cell velocities for the results using this rotational force are also compared to the system with proliferative cells restrained in the basal layer in Figure~\ref{fig:rotcelldeaths} and Figure~\ref{fig:rotcellvelocities}. 
As would be expected, given that the system maintains a consistent proliferative cell population, the velocities and cell deaths remain close to those of an equivalent system with restrained proliferative cells. 

By using the rotational force we provide a system which is able to maintain desired proliferative cell population sizes in the basal layer whilst still allowing motility of proliferative cells within the layer.
While other methods approach the problem by focusing on keeping proliferative cells in the basal layer, we approach the problem by minimising the number of differentiated cells that enter the basal layer. 
This mechanism is a robust system for simulations on the time scale of months to years using the force magnitudes trialled for this paper, and used previously in the literature. 
Though this mechanism alone is sufficient for maintaining high population densities in the model, it would likely be only one of several factors, such as membrane adhesion, contributing to proliferative cell maintenance in the system.

\section{Conclusion}
We have shown that multicellular models of epithelial tissues require the inclusion of an additional mechanism to maintain a desired density of proliferative cells in the basal layer. 
Using the base model, based on previous models in the literature (without a rotational force), we are unable to maintain proliferative cell densities in the basal layer of the IFE without a very high level of adhesion to the basal membrane. 
The steady state proliferative cell densities are particularly unstable when the spring length between the daughter cells at division is very small.
Small spring lengths at division in the model equate to a more realistic division model as cell growth starts with two essentially overlapping cells, while larger spring lengths at division add the second cell at a small translation from the other.
A decrease in the proliferative cell population will affect the dynamics and morphology of the tissue. 
Particularly, in the case of the IFE, it decreases vertical cell velocities and cell death rates. 

We showed that applying a rotational force to the differentiated daughter cell during the M phase is sufficient to maintain proliferative cell populations in the basal layer for any given spring length. 
The rotational force enforces an orientation on the cell pair during cell growth and division and hence reflects the dynamics that would be seen if directed division occurs in the epidermis.
In reality, we would expect directed division to be one of several factors ensuring that a proliferative population is maintained.
The rotational force approach also allows movement of the proliferative cells in the basal layer where other strategies, such as fixed stem cells, limit proliferative cell movement. 
Although it is not known exactly what the density of proliferative cells is in human epidermis, the inclusion of this force has enabled us to produce higher proliferative cell counts (and therefore densities), limited only by volume restrictions.
Consequently, this force can be used, in combination with other methodologies such as increased adhesion, to increase the robustness of multicellular models and help eliminate false hypotheses which could occur due to decreased basal populations and atypical cell velocities.

The rotational force concept is motivated by the observed regulation of division direction, by the mitotic spindle. 
Consequently, this methodology could be used to explore the impact of the spindle on epidermal tissue morphology, and may also be applicable in other epithelial tissue where regulation of the mitotic spindle is expected to occur.
The regulation of the mitotic spindle is potentially important to many tissues, however the sub-cellular machinery, as well as the molecular and mechanical regulators of spindle orientation, are highly complex \cite{ahringer03,lu13,mcnally13,nestor-bergmann14}. 
Future investigations will look at how the rotational force relates to the mitotic spindle regulation, and the sub-cellular mechanisms determining orientation. 
We are also interested in investigating the effect of misalignment of the spindle on tissue morphology and other potential applications for regulated spindle orientation in epithelial tissues.

Some limitations of the model presented here are the use of spherical cells and a flat basal membrane.
Though we know that cells in the lower layers are close to spherical in shape, in the upper layers cells become significantly flatter.
The use of ellipsoidal shapes in future models will allow us to investigate any potential effects the small deviation from spherical shape has on the oriented division.
The use of a flat membrane in the model is a common simplification in epidermal models \cite{kobayashi16,li13}, however it would be interesting to explore the use of an oriented force with an undulating membrane.
This raises further interesting questions for the force: would the orientation be vertical or perpendicular to the membrane?
This is a question we are interested in exploring further in future work.

Though we only consider perpendicular orientation with asymmetric division in this paper, the same concept applies for symmetric division parallel to the membrane. 
As discussed in Section~\ref{sec:modellingmitosis} there are several opposing theories on division lineage and direction in the literature, with several studies supporting the idea of an asymmetric population with some cells dividing asymmetrically and some symmetrically \cite{clayton07,sada16,simons11}.
Future work will look at incorporating both perpendicular asymmetric and parallel symmetric division directions to further investigate this hypothesis.
The inclusion of symmetric division would significantly alter the dynamics in the basal layer, but would not alter the dynamics much in the suprabasal layers assuming the proliferative population count remained roughly constant. 
It would be interesting to investigate how this proliferation rate, the balance between symmetric and asymmetric division, and division orientation could be regulated in this tissue, such as through long range signalling, using the model. 
This will help lead to a better understanding of the mechanisms by which the IFE regulates its structure, particularly its height.

\section*{Acknowledgements}	
This research was supported by an Australian Government Research Training Program (RTP) Scholarship, and in part conducted and funded by the Australian Research Council Centre of Excellence in Convergent Bio-Nano Science and Technology (project number CE140100036).

\section*{Supplementary Materials}

\begin{description}
	\item [Supplementary Figure 1] Supplementary information for Figure~\ref{fig:rotremains}. A plot of the cell loss for simulations which include a rotational force. This plot shows the results for each individual realisation and the calculated steady state population estimate ($\beta$) for data. 
	\item [Supplementary Movie 1] Supplementary information for Figure~\ref{fig:liresults}. A movie of a simulation using the base model. The red cells are proliferative cells, and the blue cells are terminally differentiated cells. Simulation output frequency is every 5 days.
	\item [Supplementary Movie 2] A movie of a simulation using the model with a rotational force included. The red cells are proliferative cells, and the blue cells are terminally differentiated cells. Parameters for the simulation are: spring length at division $l_d=10^{-3}$~CD, membrane adhesion $\alpha^*=500~\mu N$, and torsional spring constant $k_\phi = 100~\mu N$. Simulation output frequency is every 5 days.
\end{description}

\bibliography{references_paper1_journalabbrev}

\end{document}